\documentclass[conference]{IEEEtran}
\IEEEoverridecommandlockouts
\usepackage{amsmath,amssymb,amsfonts}
\usepackage{algorithmic}
\usepackage{graphicx}
\graphicspath{{figure/}}
\usepackage{textcomp}
\usepackage{xcolor}
\usepackage{comment}
\usepackage{hyperref}
\usepackage{subcaption}
\def\BibTeX{{\rm B\kern-.05em{\sc i\kern-.025em b}\kern-.08em
    T\kern-.1667em\lower.7ex\hbox{E}\kern-.125emX}}

\newcommand{\TOMOGAN}{TomoGAN}
\newcommand\wicase[2]{$W$=#1:$I$=#2}
\newcommand\wircase[3]{$W$=#1:$I$=#2:$R$=#3}

\usepackage[noadjust]{cite}

\newif\iffinal

\iffinal
    \newcommand\ian[1]{}
    \newcommand\zliu[1]{}
    \newcommand\tb[1]{}
    \newcommand\raj[1]{}
\else
    \newcommand\ian[1]{{\color{red}[ Ian: #1 ]}}
    \newcommand{\zliu}[1]{{\textcolor{blue}{ [Zhengchun: #1] }}}
    \newcommand{\tb}[1]{{\textcolor{violet}{ [Tekin: #1] }}}
    \newcommand{\raj}[1]{{\textcolor{green}{ [Raj: #1] }}}
\fi

\begin{document}

\title{Deep Learning Accelerated Light Source Experiments 
\thanks{\textsuperscript{*}Both authors contributed equally to this research.}
}

\author{\IEEEauthorblockN{Zhengchun Liu\textsuperscript{*}}
\IEEEauthorblockA{\textit{Data Science and Learning} \\
\textit{Argonne National Laboratory}\\
Lemont, IL, USA \\
zhengchun.liu@anl.gov}
\and
\IEEEauthorblockN{Tekin Bicer\textsuperscript{*}}
\IEEEauthorblockA{\textit{Data Science and Learning} \\
\textit{Argonne National Laboratory}\\
Lemont, IL, USA \\
bicer@anl.gov}
\and
\IEEEauthorblockN{Rajkumar Kettimuthu}
\IEEEauthorblockA{\textit{Data Science and Learning} \\
\textit{Argonne National Laboratory}\\
Lemont, IL, USA \\
kettimut@anl.gov}
\and
\IEEEauthorblockN{Ian Foster}
\IEEEauthorblockA{\textit{Data Science and Learning} \\
\textit{Argonne National Laboratory}\\
Lemont, IL, USA \\
foster@anl.gov}
}

\author{Zhengchun Liu\textsuperscript{*}, Tekin Bicer\textsuperscript{*}, Rajkumar Kettimuthu, Ian Foster \\
Data Science and Learning Division, Argonne National Laboratory, Lemont, IL 60439, USA \\
\{zhengchun.liu, bicer, kettimut, foster\}@anl.gov}

\maketitle



\begin{abstract}
Experimental protocols at synchrotron light sources typically process and validate data only after an experiment has completed, which can lead to undetected errors and cannot enable online steering. 
Real-time data analysis can enable both detection of, and recovery from, errors, and optimization of data acquisition. 
However, modern scientific instruments, such as detectors at synchrotron light
sources, can generate data at GBs/sec rates.
Data processing methods such as the widely used computational tomography usually require considerable computational resources, and yield poor quality reconstructions in the early stages of data acquisition when available views are sparse. 
We describe here how a deep convolutional neural network can be integrated into the real-time streaming tomography pipeline to enable better-quality images in the early stages of data acquisition. 
Compared with conventional streaming tomography processing, our method can significantly improve tomography image quality, deliver comparable images using only 32\% of the data needed for conventional streaming processing, and save 68\% experiment time for data acquisition. 
\end{abstract}

\section{Introduction}
Synchrotron light sources can provide extremely bright high-energy x-rays that
can penetrate thick materials and be focused on small regions. These x-rays can
then be used for advanced experiments, including studies of the
internal morphology of materials and samples with high
spatial (atomic and molecular scale) and temporal resolutions ($<$100 ps).
The Advanced Photon Source (APS) at Argonne National Laboratory (ANL) is an
advanced synchrotron radiation facility that hosts thousands of scientists annually
from a wide variety of communities, such as energy, materials, health, and life
sciences~\cite{aps-web, aps-science-2018}.

APS experiments can generate massive amounts of data in a short time.
For example, tomographic imaging beamlines can collect 1500 projections
(each 2048$\times$2448 pixels) in nine seconds using an Oryx
detector~\cite{flir-detector}: a rate of more than 1~GB/s.
These experiments may be performed to observe time-dependent phenomena that
spread over long time periods (weeks in some cases), resulting in large
datasets. Another example is ptychographic imaging, where current
high-performance detectors collect 3000, typically 1~MB, frames per second. Imaging a large sample, such as an integrated circuit with 1 cm\textsuperscript{2}
scanning area, can take months and generate petabytes~\cite{aps-raven}.  

These data generation rates, coupled with long experimentation times,
make it easy to generate petabytes of measurement data. 
The management of experimental data at this scale, in terms of both time and size, is challenging and requires advanced analysis techniques for timely feedback.
Methods are urgently needed that can reduce the amount of data collected, 
or permit real-time determination of whether specific data are useful.

We propose here a data analysis pipeline that uses a deep learning (DL) model to enhance the quality of reconstructed images obtained via
processing of streaming experimental data from synchrotron beamlines. 
We focus on tomography data, a common imaging modality at synchrotrons.
We show that conventional streaming tomographic reconstruction plus deep learning image enhancement 
can deliver performance that significantly surpasses that of conventional reconstruction alone, 
in terms of both image quality and throughput.

\begin{figure*}[ht]
\centering
\begin{subfigure}[b]{.323\textwidth}
\includegraphics[width=\columnwidth]{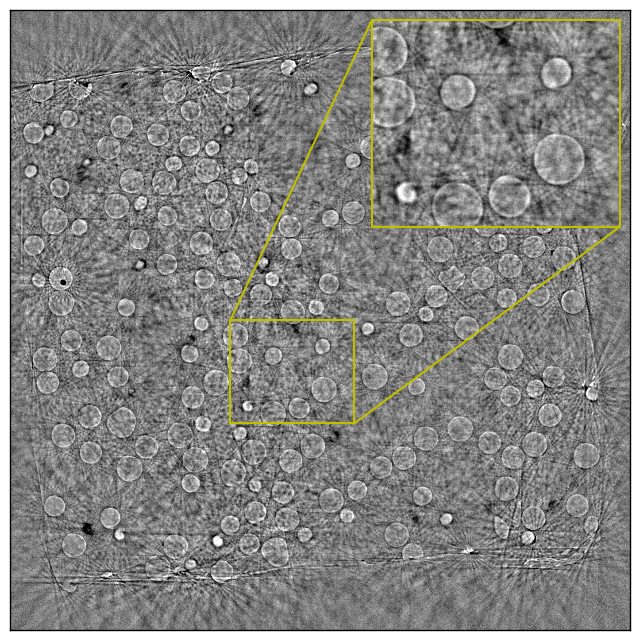}
\caption{Conventional at 462s.}
\end{subfigure}
\begin{subfigure}[b]{.323\textwidth}
\includegraphics[width=\columnwidth]{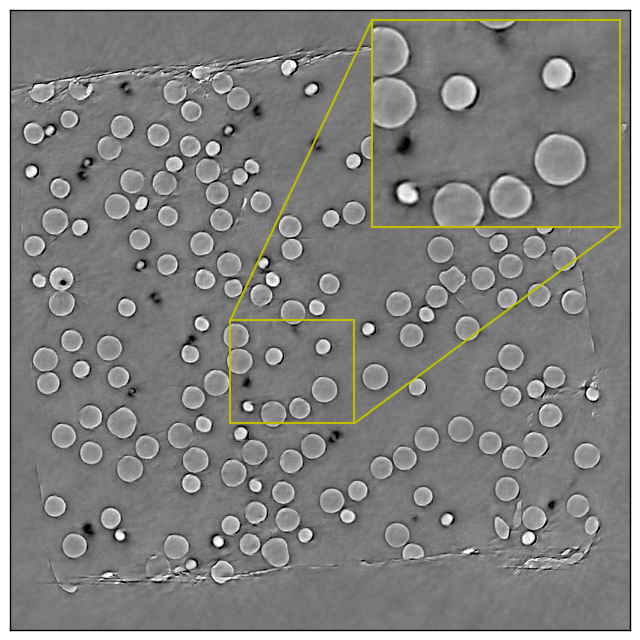}
\caption{Proposed at 462s.}
\end{subfigure}
\begin{subfigure}[b]{.323\textwidth}
\includegraphics[width=\columnwidth]{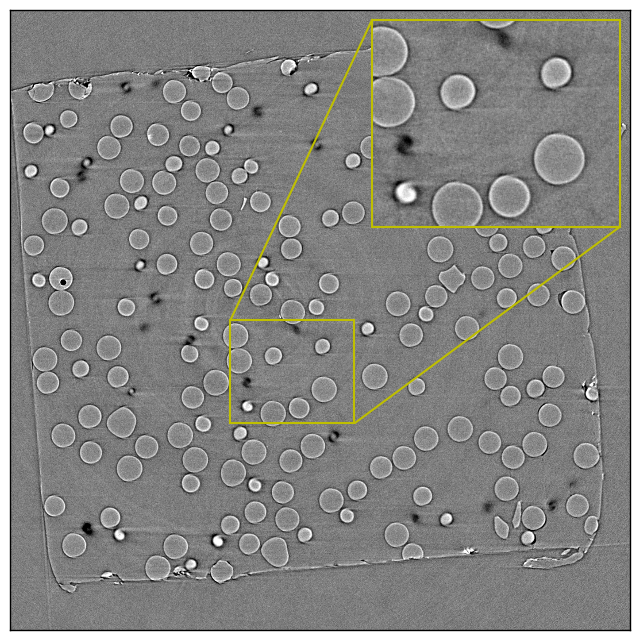}
\caption{Conventional at 1433s.}
\end{subfigure}
\caption{Streaming tomography image quality, with and without enhancement: 
(a) with data up to 462s (480 projections), before enhancement; 
(b) with the same data, after enhancement;
(c) with data up to 1433s (1504 projections), before enhancement.
}
\label{fig:teaser}
\end{figure*}

For example, \autoref{fig:teaser} shows how the integration of deep learning enhancement into the tomographic pipeline can generate images with comparable quality to those produced by conventional methods, but using only 32\% of experiment time (320 versus 1000 seconds) and after acquiring, streaming, and processing only 32\% as much data (480 versus 1504 X-ray projections).
Thus, our method can provide both three times faster turnaround time for domain scientists and three times increased throughput for the light source and computing facility. 
These improvements are also important 
as enablers of experiment steering, where quick turnaround is required.

Specifically, our paper makes the following contributions:
\begin{itemize}
\item We propose and implement a pipelined workflow for tomography reconstruction with streaming data (as shown in \autoref{fig:tomopipeline}) 
for realtime streaming analysis;
\item We repurpose and retrain \TOMOGAN{} (a GAN~\cite{2014arXiv1406.2661G}-based deep learning model originally designed for low dose X-ray tomography~\cite{liu2019tomogan}) and integrate it into our workflow for image quality enhancement (as shown in \autoref{fig:sysoverview});
\item We evaluate our system with two real-world data sets collected at APS and provide insightful analysis on performance improvements.
\end{itemize}
\section{Background}
\begin{figure*}[ht]
\includegraphics[width=\textwidth]{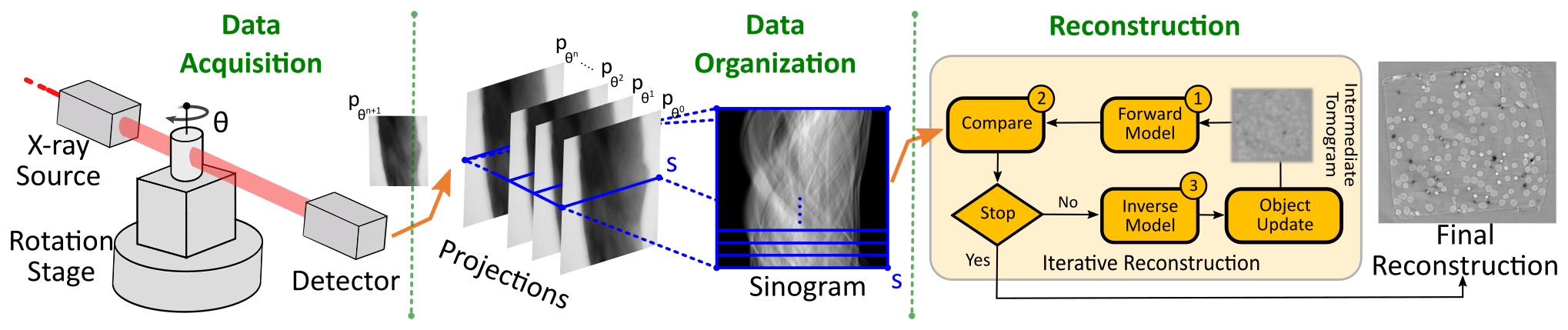}
\caption{Tomographic data acquisition and reconstruction pipeline. The steps are described in the text.}
\label{fig:tomopipeline}
\end{figure*}

We introduce the computed tomography (CT) image analysis pipeline used at synchrotron
light sources, and the use of deep learning methods for enhancing reconstructed images.
\autoref{fig:tomopipeline} illustrates the tomographic data acquisition,
management, and analysis phases at synchrotron light sources.  During the data
acquisition phase, a sample is placed on a rotation stage and illuminated by
x-ray. As x-rays pass through the sample, the photons---attenuated to a degree
determined by the thickness and density of the object---are
measured by the detector. 
The
corresponding measurement is called a {\em projection}.  A tomography experiment
collects projections from different rotations
($\theta$), with typically a fixed exposure time for each. An ideal experiment
collects projections, $P=\{P_{\theta^0}, P_{\theta^1}, \dots,
P_{\theta^n}\}$, that fully cover the sample. 

Beer's law shows the underlying mathematical model for the measurement
process~\cite{beer1852bestimmun}: 
\begin{equation}
I_\theta(s) = I_0(s) \exp\left[-p_\theta(s)\right],
\end{equation}
where $I_0(s)$ is the incident x-ray illumination on the sample and
$I_\theta(s)$ are the collected measurements at a number of $\theta$s, as a result of a tomographic scan. $p_\theta(s)$ represents a cross
section of projections (shown in blue in the central section of \autoref{fig:tomopipeline}), known as a {\em sinogram}. For parallel beam
geometry, measurements in a sinogram correspond to a cross section of the
target sample. The tomographic reconstruction process aims to recover 2D cross
section images of a sample from their corresponding sinograms.

Iterative reconstruction approaches aim to solve:
\begin{equation}\label{eq:problem}
    \hat{x}=\underset{x\in C}{\textrm{argmin}}\:\|y-Ax\|^2+R(x),
\end{equation}
where $\hat{x}$ is the reconstructed tomogram, $A$ is the forward model, $y$ is
the sinogram, $R(x)$ is a regularizer functional, $x$ is the search variable,
and $C$ is a constraint on $x$. 

Iterative approaches use statistical models to converge a solution that is
consistent with measurements. They consist of the three steps shown in the
reconstruction phase in \autoref{fig:tomopipeline}.  First, a forward model is
applied to an intermediate image estimate in order to find a measurement. Then, the
estimated and real measurements are compared. Finally, the estimated image is
updated according to the difference between the real and estimated measurements. These
steps are repeated until a user-defined constraint is met, such as total number of
iterations or error threshold.

Fast reconstruction of tomographic datasets is important to permit real-time feedback,
such as when reconstruction of a limited number of projections suffices to determine that no more data need be collected.
Although iterative reconstruction algorithms require more
computation than analytical approaches, they provide superior
reconstruction quality with {\em incomplete} or {\em limited measurements}
data.  Furthermore, many parallelization techniques have been developed to improve
their performance.  

A naive parallelization technique is
to distribute one sinogram to each process and have the processes perform independent
reconstructions in parallel. However, while this method can reconstruct many small-
to medium-scale datasets successfully, the reconstruction time for large datasets can be
long,  especially for those that require many iterations. 

Advanced parallelization
techniques, such as in-slice parallelization~\cite{bicer2017real,
bicer2016optimization} and memory-centric~\cite{hidayetoglu2019memxct}
approaches, address the limitations of the naive approach.  In-slice
parallelization replicates sinogram and image among the processes and perform
global reduction at the end of each iteration; therefore, the portions of the 
same sinogram can be reconstructed by multiple processes.  Memory-centric reconstruction, in
contrast, uses memoization and domain partitioning to split single sinogram
reconstruction to multiple processes.  Both of these advanced techniques are
suitable for quasi-real-time reconstruction of large datasets and being used at
synchrotron light source facilities~\cite{duke2016time, gursoy2015maximum}.


\section{DL-Enhanced X-Ray Image Reconstruction}
\begin{figure*}[ht]
\center
\includegraphics[width=\textwidth]{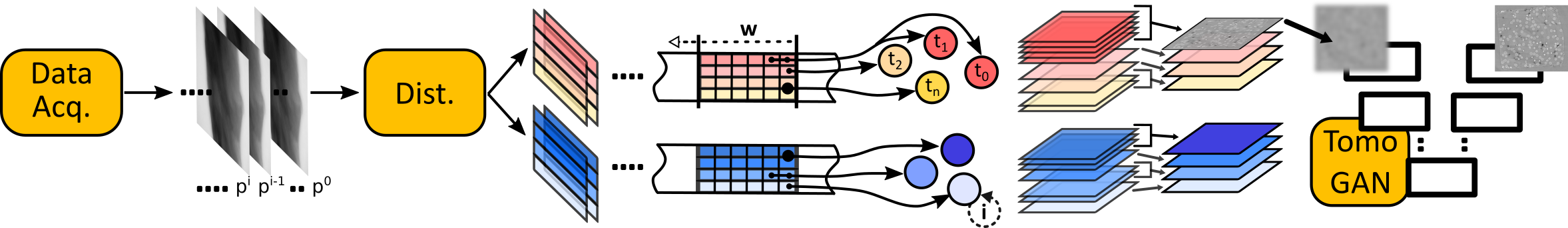}
\caption{Tomographic reconstruction on a streaming experimental data with DL denoising. $t_0, t_1, \dots, t_n$ are separate threads. }
\label{fig:sysoverview}
\end{figure*}

We next introduce our runtime system, which is optimized for
reconstruction of streaming tomography datasets, and describe its integration with
\TOMOGAN{}, our advanced GAN-based image restoration approach: see \autoref{fig:sysoverview}.

\subsection{Reconstruction of Streaming Experimental Data}
\noindent
\textbf{Data Acquisition:} As shown in \autoref{fig:sysoverview},
our system first acquires data from the tomographic experiment.  
Recall that tomographic
data acquisition is performed while a sample is rotated on a rotation axis.
Data acquisition may be fixed-angle or interleaved.  Fixed-angle
acquisition starts at a specified angle and advances
by a fixed offset until a specified final angle is reached.  For
instance, if the angle offset is $1^{\circ}$ and the experiment is set to
collect 180 projections starting from $0^{\circ}$, then the data acquisition
results in a set of projections $(p^0, p^1, \dots, p^i, \dots, p^{179})$  with
$(0^{\circ}, 1^{\circ}, \dots, 179^{\circ})$ angles.

Interleaved data acquisition also starts
at a specified angle and advances by a fixed
offset. However, acquisition proceeds in several rotations, with
each rotation starting at a different degree, resulting in {\em interleaved}
projections among rotations. For example, an interleaved data acquisition
configuration may consist of 10 rotations, each with 18 projections,
resulting in a set of projections with $(0^{\circ}, 18^{\circ}, \dots,
162^{\circ}, 1^{\circ}, 19^{\circ}, \dots, 163^{\circ}, 2^{\circ}, \dots)$
angles.

Interleaved acquisition has three advantages for real-time
reconstruction.  First, the generated projections provide full coverage with
fewer projections, and thus full volume reconstruction can start
sooner. Second, interleaved acquisition improves the convergence rate of the 
reconstruction~\cite{beekman_2001}. Third, artifacts due to a small number of
projections, e.g. dose artifacts, can be addressed with advanced iterative
reconstruction techniques and DL-enhanced denoising approaches.

\noindent
\\\textbf{Distributor:} This component receives projections
from the data acquisition component and partitions them across the
reconstruction processes.  Partitioning
is performed according to the sinograms (illustrated with colored rows in
Fig.~\ref{fig:sysoverview}). 
For instance, if a projection consists of 1024
rows and there are two reconstruction processes, then each process
receives 512 rows. This distributed-memory parallelization allows for
scaling up to the number of sinograms in the projection.

\noindent
\\\textbf{Reconstruction:} 
This component receives partitioned sinograms and performs analysis
according to user-defined configuration parameters and algorithm. The runtime
system uses a sliding window to handle streaming sinogram data.
Specifically, each partitioned projection (set of rows/sinograms) data is
pushed to a (MPI) process buffer, which is then iterated with a window.

The window size parameter, $W$, set by the user,
determines the number of projection rows at any reconstruction event. Our
system triggers reconstruction after receiving $W$ rows, and thus the projection
consumption rate can be adjusted according to available computational resources and
quality needs.

The projection data in the window are pulled by the threads associated with a process, for shared memory parallelization
and reconstruction. Threads use in-slice parallelization, where the
intermediate tomograms (or reconstructed image slices) are replicated among, and are
 independently updated, according to their corresponding
sinograms. For example, assume that the distributor assigns 512
rows/sinograms to each process, and that each sinogram is being used to
reconstruct a tomogram with dimension 1024$\times$1024. If the process has 256 threads, then the runtime system allocates 256 tomogram replicas for each
sinogram, which effectively results in total buffer size of 256$\times$512$\times$1024$\times$1024. 
This replication-based parallelization eliminates race conditions
during reconstruction, since each thread can operate on its own tomogram
replica. 

Our use of iterative reconstruction algorithms means that the number of
iterations, $I$, is another parameter that can be set by the user. 
For each triggered reconstruction event, the
threads iterate on the window data $I$ times. Thus $I$ provides
another way to adjust computational throughput and image quality. 

After each iteration, our system synchronizes (reduces) the replicated tomograms 
so that the correct tomogram can be recovered and used for the next iteration. 

\subsection{Reconstructed Image Enhancement} 
We append 
\TOMOGAN{}, an image quality enhancement model based on generative adversarial networks~\cite{2014arXiv1406.2661G} originally developed for low-dose X-ray imaging in \cite{liu2019tomogan}, to the streaming tomographic processing pipeline to enable online enhancement of image quality.
We have shown in previous work~\cite{liu2019tomogan} that, once trained on one sample, \TOMOGAN{} can be applied effectively to other similar samples, even if X-ray projections of those samples are collected at a different facility and show different noise characteristics. 

\section{Experimental Results}
We evaluate our system with respect to both the quality of the 
reconstructed images and the time required to process an image.
We work with two real-world experimental datasets collected at APS, each with different runtime configuration parameters, specifically {\em window size} and {\em iterations}.
For ease of reference, we use the notation \wircase{X}{Y}{Z} to denote streaming tomography images after \texttt{Z} rotations with a window size of \texttt{X} and with \texttt{Y} iterations performed for each update of the output image.

Our experimental datasets include \texttt{Shale} and \texttt{Glass} samples
that are imaged at APS at Argonne. \texttt{Shale} is an X-ray microtomography
dataset of a shale sample from the Upper Barnett Formation in
Texas~\cite{kanitpanyacharoen2013comparative}; it contains
tiny features (pores) with irregular shapes and sizes that are challenging to
reconstruct.  The dataset consists of 1501 projections, each of 1792$\times$2048 pixels. 
The \texttt{Glass} dataset is of a set of borosilicate glass
spheres of different sizes~\cite{singh2017varied}; it consists of 1500
projections, each of 2160$\times$2560 pixels. Both datasets are
publicly available and can be downloaded from TomoBank~\cite{de2018tomobank}.

We used the simultaneous iterative reconstruction technique~(SIRT) for tomographic reconstruction~\cite{andersen1984simultaneous}. 
We varied the window size $W$ over \{16, 32, 64, 128, 256\} and
the number of iterations $I$ over \{1, 5, 10\}. 
We simulated interleaved data acquisition, where the number of projections is set to the window size for each full rotation; thus, the window buffer is guaranteed to contain projections from angles that capture the full view of the sample for every window configuration.

The combination of different window sizes and iterations resulted in different reconstruction qualities and computational characteristics, as shown in~\autoref{tbl:cong-cmp}.  
We see that image updates can happen as frequent as once per second. 
The averaged sustained data consumption rate quantifies the number of X-ray projections per second that our workflow can process.
Thus the workflow can achieve real time data analysis if the data acquisition rate is less than the sustained rate.
\begin{table*}[htb]
\centering
\caption{Data processing time for different configurations, for \texttt{Glass} and \texttt{Shale} datasets. Refresh time is the time it takes to generate an update. The sustained data consumption rate is measured by the number of projection processed per second.}
\begin{tabular}{l|lllll|lllll|lllll}
\noalign{\hrule height 2pt}
SIRT iterations, $I$ & \multicolumn{5}{c|}{1}    & \multicolumn{5}{c|}{5}    & \multicolumn{5}{c}{10}   \\\hline
Window size, $W$  & 16 & 32 & 64 & 128 & 256 & 16 & 32 & 64 & 128 & 256 & 16 & 32 & 64 & 128 & 256 \\\noalign{\hrule height 1pt}
\texttt{Glass} Refresh time (s) & 1.5 & 1.6 & 1.8 & 2.4 & 4.0 & 7.5 & 7.9 & 9.7 & 12.9 & 20.4 & 15.4 & 16.4 & 20.1 & 26.4 & 40.8 \\\hline
\texttt{Glass} Sustained Rate (p/s) & 10.7 & 20.8 & 36.9 & 56.0 & 75.1 & 2.1 & 4.1 & 6.7 & 10.6 & 14.7 & 1.0 & 2.0 & 3.2 & 5.2 & 7.3\\\noalign{\hrule height 1pt}
\texttt{Shale} Refresh time (s) & 1.1 & 1.1 & 1.2 & 1.6 & 2.7 & 5.3 & 5.4 & 6.7 & 8.8 & 13.5 & 10.6 & 10.5 & 13.6 & 17.8 & 27.3 \\\hline
\texttt{Shale} Sustained Rate (p/s) & 15.2 & 30.2 & 52.9 & 83.0 & 112.8 & 3.1 & 6.0 & 9.8 & 15.5 & 22.2 & 1.5 & 3.1 & 4.8 & 7.7 & 11.0\\
\noalign{\hrule height 2pt}
\end{tabular}
\label{tbl:cong-cmp}
\end{table*}

We used two Argonne Leadership Computing Facility computer systems in this work:
Theta for reconstructions 
and Cooley to train \TOMOGAN{} and to run \TOMOGAN{} on the reconstructed images.
Theta consists of 4392 Intel Xeon Phi (KNL) nodes and has a peak speed of 11.60 petaflops. 
Cooley has 126 compute nodes, each with a Tesla K80 dual GPU card with 24GB memory. 

\subsection{Reconstruction Quality Improvement}
Due to dataset limitations, we split our samples, each with 1024 images of 2560$\times$2560 pixels, into 128 (12.5\% of total) for training and the rest for testing. 
We trained the \TOMOGAN{} model with 128 tomography images obtained with the configuration \wircase{32}{1}{5} and their corresponding ground truth. The ground truth we used to train \TOMOGAN{} is an offline reconstruction using SIRT with 100 iterations and all projections, i.e., the best reconstruction we can get with the state-of-the-art method. We appended the trained model to the streaming pipeline to enhance tomography images (as shown in \autoref{fig:sysoverview}) for \emph{all other} experiment configurations, i.e., every other combination of $W$ and $I$ for each updating/rotation ($R$).

The structural similarity index metric (SSIM)~\cite{ssim} is a commonly used method for measuring the similarity between two images. 
SSIM is a full reference metric; in other words, its measurement of image quality is based on an initial uncompressed or distortion-free image as reference. 
SSIM is designed to improve on traditional methods such as peak signal-to-noise ratio and mean squared error.
It ranges from 0 to 1, where 0 means two images are completely different and 1 means the two images are identical.
For example, Baker et al.~\cite{img-simi-cmp} compared ten different similarity metric for image and indicated that the SSIM metric performs the best.
In this paper, we use the SSIM between ground truth (i.e., the best possible) and the target image to quantify image quality. 
Thus a larger SSIM value means more similar to the best possible image, and thus higher image quality. 

\begin{figure*}[ht]
\center
\includegraphics[width=1.5\columnwidth]{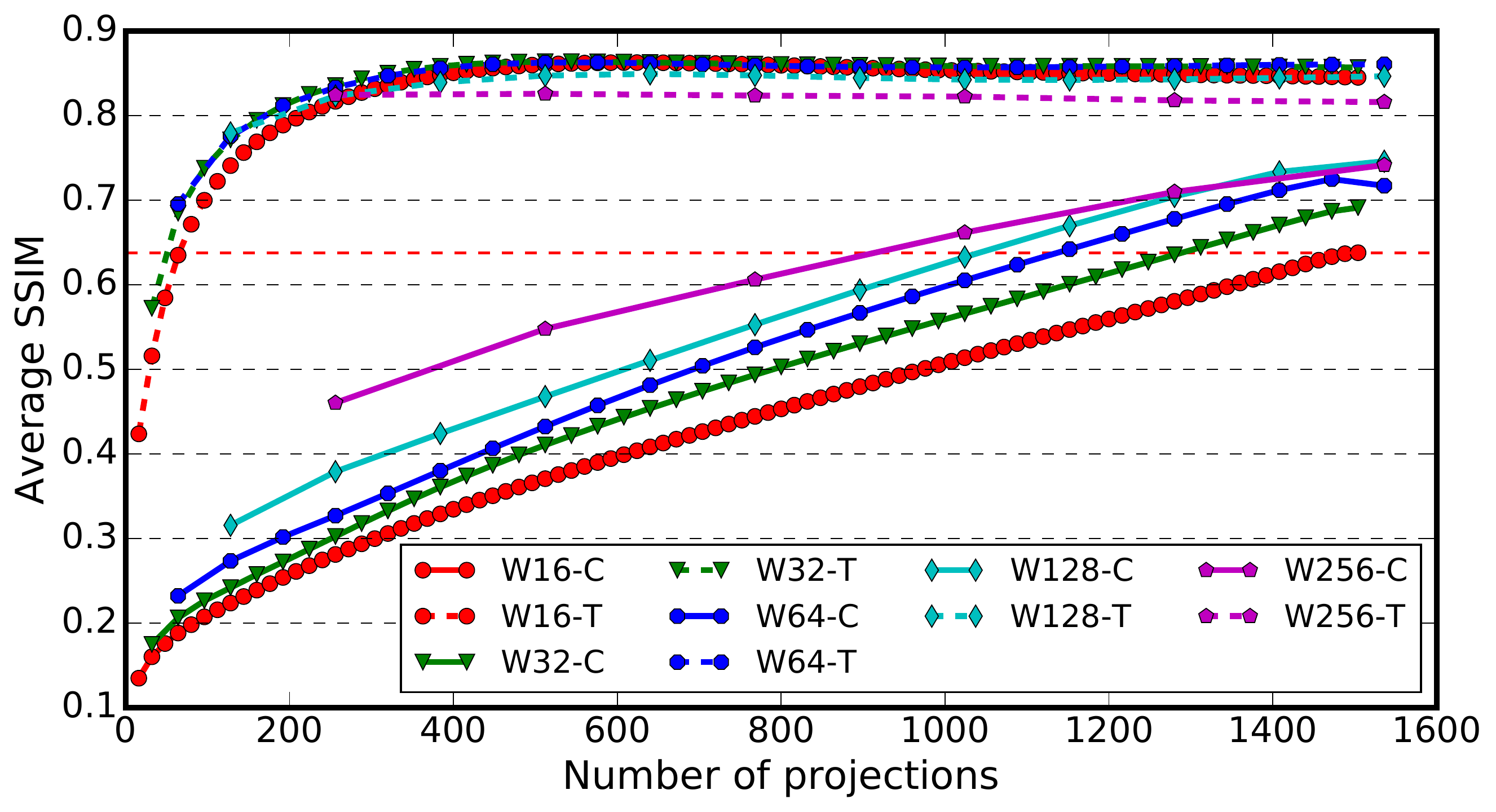}
\caption{Streaming tomography image quality improvements for \texttt{Glass}, as measured by SSIM, averaged across all slices. The labels in the legend are coded as follows: \textbf{W} is window size; \textbf{C} denotes conventional reconstruction and \textbf{T} denotes conventional plus \TOMOGAN{} enhancement. Streaming tomography processing uses 10 iterations in each case. The red dashed horizontal line shows the best result obtained with the conventional method and a window size of 16.}
\label{fig:glass-ssim-imprv-i01}
\end{figure*}

For each update (i.e., one full rotation of data acquisition) of the \texttt{Glass} sample, \autoref{fig:glass-ssim-imprv-i01} shows the image quality improvement as measured by SSIM when performing 10 SIRT iterations per update, and for different window sizes.
Each dot in \autoref{fig:glass-ssim-imprv-i01} represents the average SSIM of all tomography slices for that number of updates.
There is an update after every $W$ X-ray projections acquired, streamed, and processed. 
The timestamp of each update is proportional to the corresponding refresh time, as shown in \autoref{tbl:cong-cmp}.

\begin{figure*}[ht]
\center
\includegraphics[width=1.7\columnwidth]{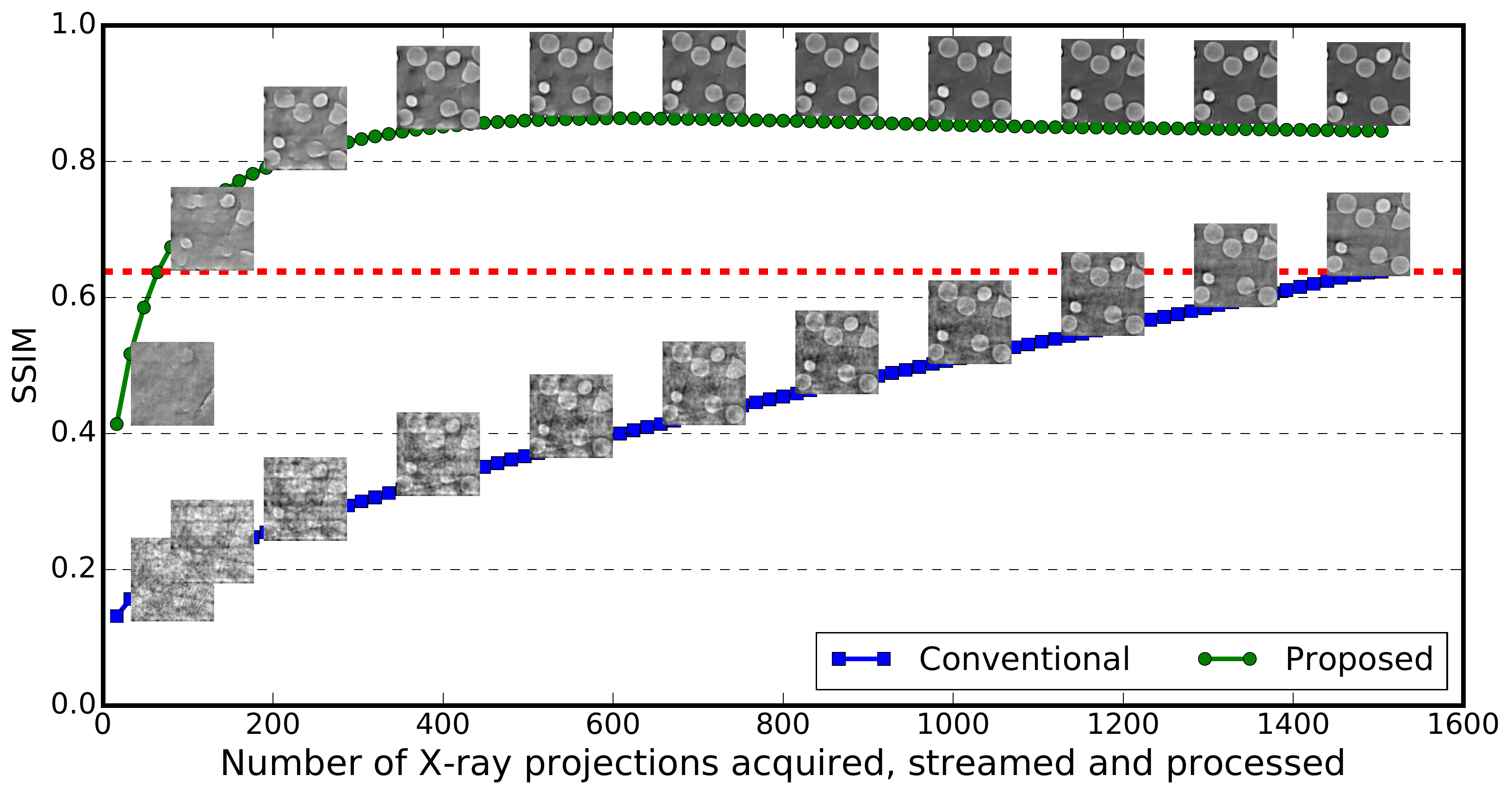}
\caption{Comparison of SSIM values, for a representative region with the conventional and proposed methods at \wicase{16}{1}, shows the improvements obtained with the latter. 
The red dashed line shows the best result obtained with the conventional method.
}
\label{fig:preview-cmp-w16-i1-1}
\end{figure*}

More specifically for the case \wicase{16}{1} (one with the most frequent updates), \autoref{fig:preview-cmp-w16-i1-1} shows the SSIM comparison with conventional streaming as well as a regional preview for every 10 updates.
We see that the image generated with \TOMOGAN{} in processing pipeline becomes, after 20 rotations (i.e., 320 projections acquired), visually comparable with the best image that the conventional streaming tomography can get at the end. 

\subsection{End-to-end Performance Evaluation}
\begin{figure*}[ht]
\center
\includegraphics[width=\textwidth]{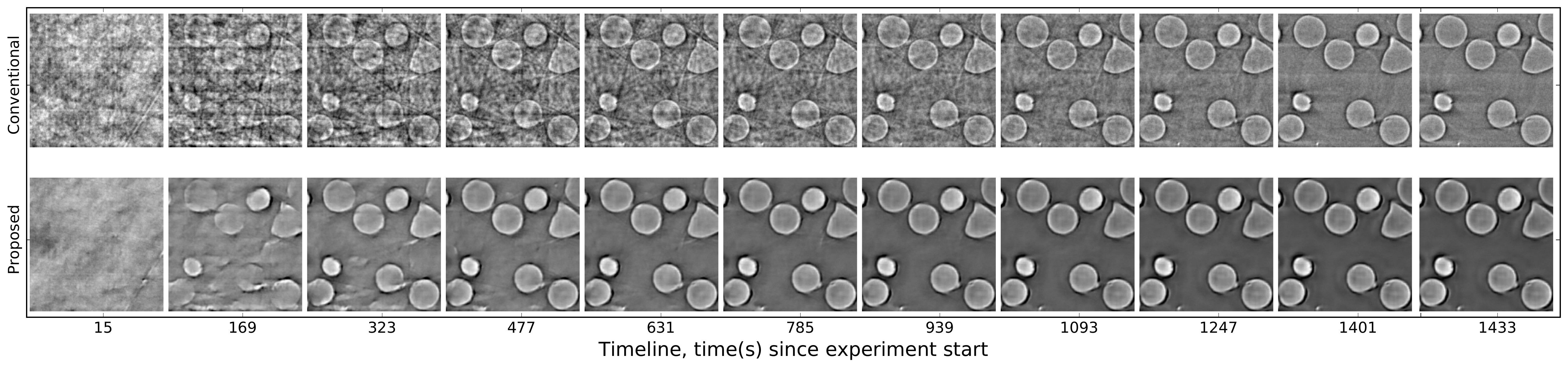}
\caption{Reconstucted images obtained as a experiment proceeds for a representative region of the \texttt{Glass} dataset. 
Image quality with the addition of deep learning (below) is significantly improved relative to conventional reconstruction alone (above).}
\label{fig:preview-cmp-w16-i10}
\end{figure*}

If we purely measure the image quality with SSIM, as shown in \autoref{fig:glass-ssim-imprv-i01}, then for \wicase{16}{10}, the best SSIM that conventional streaming tomography can achieve is 0.638.
When using \TOMOGAN{}, in contrast, the SSIM exceeds 0.638 (as shown by the horizontal red dotted line in \autoref{fig:glass-ssim-imprv-i01}) after just four rotations.
However, as shown in \autoref{fig:preview-cmp-w16-i10}, the (visual) image quality even after 11 rotations 
(i.e., at 169s, because each update takes 15.4s, as shown in \autoref{tbl:cong-cmp}) is poor. 
We thus conclude that we cannot rely only on SSIM for image quality measurement to estimate the end-to-end performance improvement. 

As an alternative, we use the naked eye to subjectively evaluate image quality to estimate the end-to-end speedup of throughput. 
We evaluate the image quality based on two factors: (1) Similarity to the best possible image quality (i.e., compare \TOMOGAN{} denoised image  with the best image from conventional streaming tomography. The best image from conventional streaming tomography is the one that is obtained after processing all the projections - for example, the one on the blue curve at 1500 projections in \autoref{fig:preview-cmp-w16-i1-1}.) and, (2) Clarity of 
features in the image (i.e., compare \TOMOGAN{} denoised image with the image from conventional streaming that is just clear enough to see all features). 
We observe: (1) as shown in \autoref{fig:preview-cmp-w16-i10}, the \TOMOGAN{} denoised image at 477s (i.e., the 31st update) is visually comparable with the best image (at 1433s) from conventional streaming: a speedup of about 3; and (2) 
features in the image using conventional streaming tomography are observable only at 477s whereas features in the \TOMOGAN{} denoised image is observable at 169s (i.e., the 11th update), again a speed-up of nearly 3. 

\begin{figure*}[ht]
\center
\includegraphics[width=\textwidth]{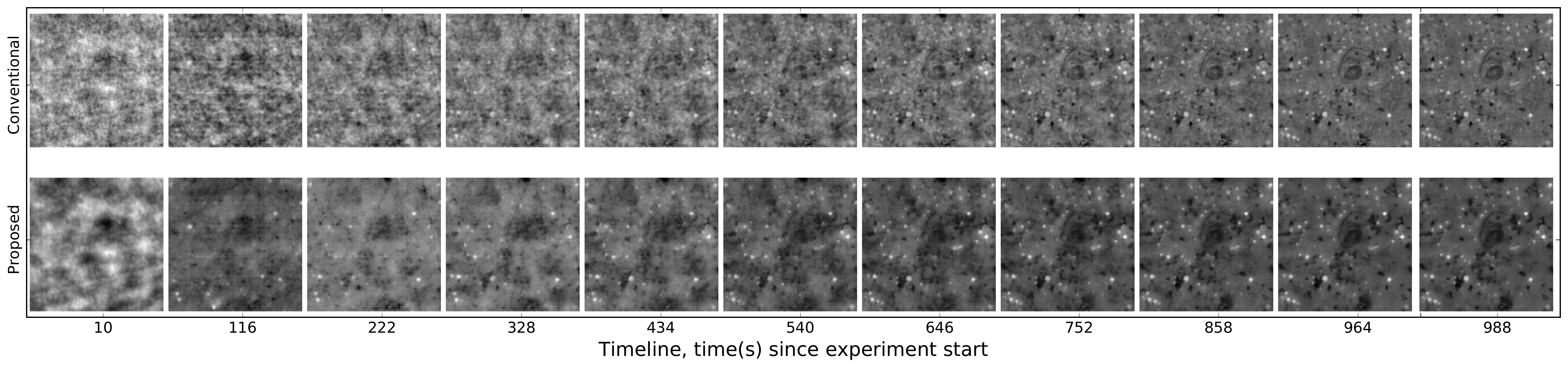}
\caption{Conventional recontruction (above) vs.\ conventional plus deep learning (below), as in Fig.~\ref{fig:preview-cmp-w16-i10}, but for the \texttt{Shale} dataset.}
\label{fig:preview-cmp-w16-i5-shale76}
\end{figure*}

We also evaluated our method using the \texttt{Shale} sample. 
The image quality improvement is demonstrated in \autoref{fig:preview-cmp-w16-i5-shale76}.
The \texttt{Shale} sample has much more high-frequency content when compare with \texttt{Glass}. 
Although the image quality improvement is still clear, the end-to-end speedup is not as good as it for \texttt{Glass} sample. The end-to-end speed up is 2x (as opposed to 3x for \texttt{Glass}) by naked eye evaluation.

\subsection{Overhead analysis}
\TOMOGAN{} takes about 290ms to process one 2560$\times$2560 pixel image in our experiments on one NVIDIA Tesla V100 GPU card. 
 \TOMOGAN{} and the tomographic reconstruction algorithm can run in parallel and \TOMOGAN{} takes significantly less time than the reconstruction algorithm. Thus, the tomographic reconstruction algorithm and \TOMOGAN{} can be effectively pipelined such that the total overhead of \TOMOGAN{} is only 290ms irrespective of the number of reconstruction steps (or the number of updates done to the output image, which is as the number of rotations performed on the sample in our experimental setup). In other words, the first update of the output image is delayed by 290ms with \TOMOGAN{} in the processing pipeline but the frequency of the subsequent updates remains the same as that of the processing pipeline. Given that \TOMOGAN{} cuts down the number of updates (and the amount of projections need to be collected) by a factor of 3 or 2, a 290ms delay in getting the first update is negligible.  
Therefore, \TOMOGAN{} also does not affect the sustained data consumption rate, i.e., projections per seconds.
However, the limitation here is that there needs to be one GPU card per node to run \TOMOGAN{} to achieve such a low delay.
The delay will increase if \TOMOGAN{} is run on another server because of data movement latency. 
The sustained data consumption rate in \autoref{tbl:cong-cmp}, measured by the number of projection processed per second, when compared with data acquisition rate, can be used to quantify the real-time processing capability for different configurations.
\section{Related Work}
Tomographic reconstruction techniques can broadly be categorized into two
groups: {\em analytical} and {\em iterative} approaches.  Analytical
reconstruction techniques perform single-pass over the dataset, and
known to be computationally efficient compared to iterative methods; however
they are prone to measurement errors and require sufficient amount of data for
good reconstruction~\cite{Beister:Iterative_Recon_Methods:PM12}.  In contrast,
iterative approaches are resilient to noisy measurements and can provide
reasonable reconstructions even with limited data. Although iterative
reconstruction techniques require significant compute resources, their parallel
implementations enable their usage on challenging
datasets~\cite{agulleiro2011fast, treibig2012pushing,
johnson:data_parallel_iter:MPC99, jones:hybrid_recon:NC06}. Further, since
they can provide reasonable reconstructions with limited data, they are 
suitable for fast feedback workflows with rapid reconstructions.

Iterative reconstruction techniques have been successfully used to provide high
quality images, especially in medical imaging
area~\cite{Chou:Medical_Imaging_GPUs:MP11, pratx:accurate_recon_gpu:MI09,
Lee:Neuro_Imaging_CUDA:CMPB12}, where limiting dose exposure  is
important~\cite{sidky2006accurate, Jang:MultiGPU_Iterative_Recon:BI09}.  The
computational requirements of these algorithms have typically been met with
many-core architectures, such as GPUs and KNLs~\cite{astra, sabne2017model,
Stone:MRI_Recon_GPUs:JPDC08, xu:Tomo_Recon_GPUs:NC05}. Most of these works
consider the availability of all data and are not optimized for real-time
reconstruction.

Deep learning (DL) approaches have been used successfully in many scientific
imaging problems, such as denoising, feature segmentation, image restoration
and super resolution~\cite{ARAUJO201913, dula:ML:2017, 2017arXiv171110925U,
srgan, pix2pix2017}. Among these, denoising reconstructed images has been an
active area~\cite{wang2005sinogram, mohan:2014, ma2011low}. Many DL approaches
have been developed and applied to denoise reconstructed images~\cite{8332971,
Lovric:xk5008, wolterink2017generative, deep-img, jimaging4110128,
2016arXiv160806993H}. Pelt et al.\ also used a mixed-scale convolutional
neural network to improve noise in CT images, with  impressive
results~\cite{jimaging4110128, 2016arXiv160806993H}. Yang et al.~\cite{Yang2018},
use a convolutional neural network (CNN) to denoize reconstructed
images and show 10-fold improvement on signal-to-noise ratio.  In our work, we
apply our denoising method, TomoGAN~\cite{liu2019tomogan}, to streaming reconstructions and evaluate
its impact on image quality and end-to-end performance.

Real-time experimental data analysis~\cite{vogelgesang2012ufo, xu2007real,
van2015astra} and steering have been active research
areas~\cite{blaiszik2019data, wang2001high}. 
ASTRA is a popular GPU-based toolkit for processing and reconstruction of x-ray
data~\cite{van2015astra}.  UFO is another image processing
framework for synchrotron dataset that uses GPUs for fast feedback and
visualization~\cite{vogelgesang2012ufo}. MemXCT is a highly optimized 
reconstruction engine for large-scale tomography datasets~\cite{hidayetoglu2019memxct}. 
In this work, we extended our efficient stream reconstruction data analysis
pipeline~\cite{bicer2017real, bicer2017trace, bicer2015rapid} with denoising
capabilities\cite{liu2019tomogan,xloop19}.

\section{Conclusions and Future work}
We presented a new method for real-time computed tomography at synchrotron light sources.
In this new method, a deep learning model is used to improve the quality of tomographic
reconstructions as data is collected, thus producing high-quality output more quickly or, alternatively, reducing the amount of data that must be collected. 

Our experimental evaluations, using real-world datasets, show significant improvement in tomography image quality and system throughput. 
In particular, the proposed method need only a fraction (as low as 1/3) 
of the data required for conventional reconstruction methods, 
thus saving not only precious beamline time but also the network and computing resources that would otherwise be required to process the data.
Thus, end-to-end experimental throughput is as much as three times greater than that of state-of-the-art conventional methods.

Much of our work can be reused 
for other synchrotron light source analysis tasks. 
For example, the data acquisition component can be used for any pixelated detector, and many modalities can be implemented by using our parallel processing framework, including correlation analysis for x-ray photon spectroscopy, ptychographic reconstruction, and fitting of fluorescence data.

In future work, we plan to explore how these methods can be integrated into an experiment steering framework, to help domain scientists correct or terminate unwanted data collection. 
We also intend to explore architecture, methods, and algorithms needed to support autonomous experiments.  

\section*{Acknowledgment}
This work was supported in part by the U.S. Department of Energy, Office of Science, Advanced Scientific Computing Research, under Contract DE-AC02-06CH11357.  
We 
acknowledge computing resources provided and operated by Argonne's Joint Laboratory for System Evaluation and Argonne Leadership Computing Facility.

\bibliographystyle{ieeetr}
\bibliography{Bibs/bicer,Bibs/denoising,Bibs/recon,Bibs/main}

\section*{License}
\noindent
The submitted manuscript has been created by UChicago Argonne, LLC, Operator of Argonne National Laboratory (``Argonne"). Argonne, a U.S. Department of Energy Office of Science laboratory, is operated under Contract No. DE-AC02-06CH11357. The U.S. Government retains for itself, and others acting on its behalf, a paid-up nonexclusive, irrevocable worldwide license in said article to reproduce, prepare derivative works, distribute copies to the public, and perform publicly and display publicly, by or on behalf of the Government. The Department of Energy will provide public access to these results of federally sponsored research in accordance with the DOE Public Access Plan. http://energy.gov/downloads/doe-public-access-plan.

\end{document}